\documentclass[12pt]{article}
\pdfoutput=1
\usepackage{putex}
\usepackage{textcomp}
\usepackage{color}
\usepackage{autobreak}
\usepackage{indentfirst}
\usepackage{graphicx}
\usepackage{float}
\graphicspath{{plots/}}
\usepackage{tabularx}
\usepackage{caption}
\usepackage{tikz, pgfplots}
\usetikzlibrary{arrows.meta, calc, positioning}
\usetikzlibrary{decorations.pathmorphing}
\usepackage{mathtools,amsfonts,amssymb}
\numberwithin{equation}{section}
\usepackage{cancel}
\usepackage{array}
\usepackage{mathrsfs}
\usepackage{subcaption}
\usepackage{epigraph}
\usepackage{epstopdf}
\usepackage{bbold}
\usepackage{enumerate}
\usepackage{cite}
\usepackage{youngtab}
\usepackage{tensor}
\usepackage{slashed}
\usepackage[aligntableaux=center]{ytableau}
\usepackage{CJKutf8}
\usepackage[utf8]{inputenc}
\usepackage[T1]{fontenc}
\usepackage{rotating}
\usepackage{multirow}
\usepackage[
      colorlinks=true,
      linkcolor=red,
      urlcolor=red,
      filecolor=black,
      citecolor=red,
      ]{hyperref}
\usepackage[noabbrev]{cleveref}

\makeatletter
\DeclareFontFamily{OMX}{MnSymbolE}{}
\DeclareSymbolFont{MnLargeSymbols}{OMX}{MnSymbolE}{m}{n}
\SetSymbolFont{MnLargeSymbols}{bold}{OMX}{MnSymbolE}{b}{n}
\DeclareFontShape{OMX}{MnSymbolE}{m}{n}{
    <-6>  MnSymbolE5
   <6-7>  MnSymbolE6
   <7-8>  MnSymbolE7
   <8-9>  MnSymbolE8
   <9-10> MnSymbolE9
  <10-12> MnSymbolE10
  <12->   MnSymbolE12
}{}
\DeclareFontShape{OMX}{MnSymbolE}{b}{n}{
    <-6>  MnSymbolE-Bold5
   <6-7>  MnSymbolE-Bold6
   <7-8>  MnSymbolE-Bold7
   <8-9>  MnSymbolE-Bold8
   <9-10> MnSymbolE-Bold9
  <10-12> MnSymbolE-Bold10
  <12->   MnSymbolE-Bold12
}{}
\let\llangle\@undefined
\let\rrangle\@undefined
\DeclareMathDelimiter{\llangle}{\mathopen}%
                     {MnLargeSymbols}{'164}{MnLargeSymbols}{'164}
\DeclareMathDelimiter{\rrangle}{\mathclose}%
                     {MnLargeSymbols}{'171}{MnLargeSymbols}{'171}
\makeatother
\begin{document}
%%%%%%%%%%%%%%%%%%%%%%%%%%%%%%%%%%%%%%%%%%%%%%%%%%%%%%%%%%%%%%%%%%%%%%%%%%%%%%%%%%%%%%%%%%%%%%
\institution{galicia}{Instituto Galego de F{\'i}sica de Altas Enerx{\'i}as (IGFAE) and Departamento de F{\'i}sica de Part{\'i}culas, \cr 
Universidade de Santiago de Compostela}
%%%%%%%%%%%%%%%%%%%%%%%%%%%%%%%%%%%%%%%%%%%%%%%%%%%%%%%%%%%%%%%%%%%%%%%%%%%%%%%%%%%%%%%%%%%%%%
\title{
Chiral algebra correlators of the \texorpdfstring{$6$}{6}d, \texorpdfstring{$\mathcal{N}=(2,0)$}{N=(2,0)} theory with a defect
}
%%%%%%%%%%%%%%%%%%%%%%%%%%%%%%%%%%%%%%%%%%%%%%%%%%%%%%%%%%%%%%%%%%%%%%%%%%%%%%%%%%%%%%%%%%%%%%
\authors{
Konstantinos C. Rigatos\worksat{\galicia}${}^{,}$\footnote{{\hypersetup{urlcolor=black}\href{mailto:konstanstinos.rigatos@usc.es}{konstanstinos.rigatos@usc.es}}}
}
%%%%%%%%%%%%%%%%%%%%%%%%%%%%%%%%%%%%%%%%%%%%%%%%%%%%%%%%%%%%%%%%%%%%%%%%%%%%%%%%%%%%%%%%%%%%%%
\abstract{  
We revisit the computation of $2$-point correlation functions of $\tfrac{1}{2}$-BPS operators in the $6$d, $\mathcal{N}=(2,0)$ theory in the presence of a surface defect. We focus on the protected sector described by the chiral algebra and reproduce its correlators using a bootstrap approach. A unique feature of this calculation is that we can completely determine the answer by using bulk-channel data only. We, also, perform the operator expansion in the defect channel which provide new predictions for the theory.
}
\date{\today}
%%%%%%%%%%%%%%%%%%%%%%%%%%%%%%%%%%%%%%%%%%%%%%%%%%%%%%%%%%%%%%%%%%%%%%%%%%%%%%%%%%%%%%%%%%%%%%
\maketitle
%%%%%%%%%%%%%%%%%%%%%%%%%%%%%%%%%%%%%%%%%%%%%%%%%%%%%%%%%%%%%%%%%%%%%%%%%%%%%%%%%%%%%%%%%%%%%%
{
\hypersetup{linkcolor=black}
\tableofcontents
}
%%%%%%%%%%%%%%%%%%%%%%%%%%%%%%%%%%%%%%%%%%%%%%%%%%%%%%%%%%%%%%%%%%%%%%%%%%%%%%%%%%%%%%%%%%%%%%
\newpage
%%%%%%%%%%%%%%%%%%%%%%%%%%%%%%%%%%%%%%%%%%%%%%%%%%%%%%%%%%%%%%%%%%%%%%%%%%%%%%%%%%%%%%%%%%%%%%
\section{Prologue}
A natural way to examine a quantum field theory is to probe it in the presence of defects. For conformal field theories (CFTs) it is equivalent to introducing a new set of numbers to the CFT data. This is one reason to study correlators of local operators in the presence of the defect; it is the most feasible approach to compute the new data. These correlators have a central role in the bootstrap program for CFTs with boundaries (BCFT) and defects (dCFT) \cite{Liendo:2012hy,Billo:2016cpy,Lemos:2017vnx,Liendo:2019jpu,Bissi:2018mcq,Mazac:2018biw,Barrat:2022psm,Bianchi:2022ppi}, withal it is imperative to be able to compute correlators with efficacy since they are the most basic observables of any theory. Most studies have been at weak-coupling, since there is a plethora of standard methods leading to concrete answers \cite{Cuomo:2021kfm,Cuomo:2022xgw,Giombi:2021uae,Giombi:2020rmc,Bianchi:2022sbz,Gimenez-Grau:2022ebb,Raviv-Moshe:2023yvq,SoderbergRousu:2023pbe,SoderbergRousu:2023nvd,Trepanier:2023tvb,Giombi:2023dqs,Baerman:2024tql,Chalabi:2025nbg,deLeeuw:2023wjq,Georgiou:2023yak,Linardopoulos:2025ypq,Linardopoulos:2025mav}. At strong coupling, where the AdS/CFT provides us with a trustworthy dual description, our understanding is not that advanced and results were scarce and unsystematic \cite{Chiodaroli:2016jod,Giombi:2017cqn,Wang:2020seq,Drukker:2020swu,Giombi:2023zte}. 

This is to be contrasted with the dizzying array of results for correlators of local operators in CFTs without boundaries/defects at strong coupling. The bootstrap approach, initiated in \cite{Rastelli:2016nze,Rastelli:2017udc} for holographic $4$-point functions, has been remarkably successful eschewing the complications due to explicit input of couplings by relying on consistency conditions. It has been extended to correlators of higher points, beyond tree-level, different set-ups, and even beyond the supergravity (SUGRA) regime, see \cite{Bissi:2022mrs} for a recent review. Related is the recent advancement of employing this program to holographic defect and defect-like theories \cite{Ferrero:2021bsb,Barrat:2021yvp,Meneghelli:2022gps,Chen:2023oax,Gimenez-Grau:2023fcy,Ferrero:2023znz,Ferrero:2023gnu,Chen:2023yvw,Zhou:2024ekb,Chen:2024orp,Chen:2025yxg}. This sets the stage to a more systematic understanding of calculating correlators in these set-ups. Here we make further progress to that direction for the $6$d, $\mathcal{N}=(2,0)$; a non-Lagrangian strongly-coupled theory. While lacking a conventional description is an objective hindrance in computing observables, we can use the AdS/CFT to make headway. At large-N it is dual to the $\text{AdS}_7 \times \text{S}^4$ solution of $11$d SUGRA, with the $\tfrac{1}{2}$-BPS local operators mapped to an infinite Kaluza-Klein (KK) tower of scalars and the $\tfrac{1}{2}$-BPS defect being a probe M$2$-brane lying flat along an $\text{AdS}_3 \subset \text{AdS}_7 \times \text{S}^4$. 

Additionally, there is a property that not only the $\mathcal{N}=(2,0)$ theory possesses, but all maximal superconformal field theories (SCFTs) share; the existence of a protected sub-sector of operators which closes on itself \cite{Beem:2013sza,Beem:2014kka,Chester:2014mea,Beem:2016cbd}. To form the operators of that sub-sector we restrict a class of operators of the theory ($\tfrac{1}{2}$-BPS operators) to a lower-dimensional sub-space (two-dimensional plane) and perform a certain ``twist'' by restricting the R-symmetry polarizations to special values. Crucially this sub-sector contains only operators of short multiplets and is independent of marginal deformations in case there are any \cite{Beem:2013sza}. In $6$d this assumes the form of a unitary chiral algebra \cite{Beem:2014kka} and imposes strong constraints on the correlators of the theory. Non-trivial unitarity bounds are captured analytically \cite{Beem:2014kka,Liendo:2015ofa,Lemos:2015orc,Beem:2018duj} and the chiral algebra data can, also, be used to augment the power of the numerical bootstrap \cite{Chester:2014mea,Beem:2013qxa,Beem:2014zpa,Beem:2015aoa,Lemos:2016xke,Beem:2016wfs,Gimenez-Grau:2020jrx,Bissi:2020jve}. The chiral algebra of the $\mathcal{N}=(2,0)$ theory has been used to extract protected CFT data and spectrum information for $\tfrac{1}{2}$-BPS operators \cite{Beem:2014kka,Chester:2018dga}, with \cite{Behan:2021pzk} providing further evidence for the chiral algebra conjecture by explicit calculations both in the SUGRA and field theory sides and finding perfect agreement\footnote{For bootstrap constraints on the $4$-point correlators of the $\mathcal{N}=(2,0)$ theory see \cite{Arutyunov:2002ff,Heslop:2004du,Beem:2015aoa,Rastelli:2017ymc,Zhou:2017zaw,Heslop:2017sco,Abl:2019jhh,Alday:2020lbp,Lemos:2021azv,Kantor:2022epi} for analytic and numerical results.}. 

The R-symmetry ``twist'' makes the correlators of the theory meromorphic functions. This means that we need to pass to the cohomology of an appropriately chosen supercharge. Chiral correlators can, then, be computed directly in the field theory using the holomorphic bootstrap \cite{Headrick:2015gba}. The basic idea is that the meromorphic functions can be determined by the singularities dictated by the chiral algebra OPE. This approach has been used for the $\mathcal{N}=(2,0)$ theory to compute correlators in the defect-free case \cite{Rastelli:2017ymc,Behan:2021pzk}\footnote{After we submitted the first version of this work on the arXiv, we became aware of \cite{Woolley:2025qbd}, where the author performs a chiral algebra bootstrap for the $6$d, $\mathcal{N}=(2,0)$ theory without any defects, extending the results of \cite{Rastelli:2017ymc} to general charges.} and in the presence of a defect for the lowest-lying $\tfrac{1}{2}$-BPS operators \cite{Meneghelli:2022gps}.

In this work we continue in a natural way extending the results of \cite{Meneghelli:2022gps} to account for higher-KK modes. We consider $2$-point correlators of $\tfrac{1}{2}$-BPS operators in the $6$d, $\mathcal{N}=(2,0)$ theory in the presence of a $\tfrac{1}{2}$-BPS surface defect. We focus on the protected sub-sector of the theory described by the chiral algebra and bootstrap its correlation functions using the strategy outlined above. A unique feature in our case is that the defect correlator has an in-built crossing symmetry that we exploit and this allows us to completely fix these correlators by using data only from the bulk channel. Our work is, also, supplementary to the SUGRA computation of \cite{Chen:2023yvw} since it provides a highly non-trivial check of their result.
%%%%%%%%%%%%%%%%%%%%%%%%%%%%%%%%%%%%%%%%%%%%%%%%%%%%%%%%%%%%%%%%%%%%%%%%%%%%%%%%%%%%%%%%%%%%%%
\section{Preliminaries}\label{sec: prems}
The $2$-point correlators of bulk operators admit the following large-N expansion \cite{Chen:2023yvw,Chen:2024orp}:
\begin{equation}\label{eq: expansion}
\llangle S_{k_1} S_{k_2} \rrangle
= 
\underbrace{\langle S_{k_1} S_{k_2}\rangle_{\text{AdS}}+\frac{1}{N}\llangle S_{k_1} \rrangle \llangle S_{k_2} \rrangle}_{\text{disconnected contributions}}
+
\frac{1}{N^2}\llangle S_{k_1} S_{k_2} \rrangle_{\text{ tree}}+\ldots
\,      ,
\end{equation}
where in the above $\llangle \rrangle$ denotes the presence of the non-trivial defect. The disconnected contribution contains two terms. The first is given by the free propagator in AdS, while the second is the product of $1$-point functions. The first non-trivial contribution is the connected part, $\llangle S_{k_1} S_{k_2} \rrangle_{\text{tree}}$, while $\ldots$ represents higher-loop corrections which are not relevant for us. Our focus is on the non-trivial tree-level contribution. 
%%%%%%%%%%%%%%%%%%%%%%%%%%%%%%%%%%%%%%%%%%%%%%%%%%%%%%%%%%%%%%%%%%%%%%%%%%%%%%%%%%%%%%%%%%%%%%
\subsection{Superconformal kinematics and chiral algebra}
Our discussion on the basic kinematics follows closely the notation of \cite{Chen:2023yvw}. A more in-depth presentation can be found in \cite{Meneghelli:2022gps}. 

We consider $\tfrac{1}{2}$-BPS operators denoted by $S_k$. These are the protected superconformal primaries of the $6$d, $\mathcal{N}=(2,0)$ theory with scaling dimensions $\Delta_{S_k} = 2k$. They transform in the $[k,0]$ representation of the R-symmetry, described by an $\mathfrak{so}(5)_{\text{R}}$ algebra, which is the rank-$k$ symmetric traceless. It is convenient to contract the indices with an auxiliary polarization vector, $u^I$, to keep a track of the R-symmetry:
\begin{align}
S_k(x,u) = S_{I_1 \ldots I_k} u^{I_1} \ldots u^{I_k} 
\,         , 
\end{align}
and we note that the polarization vector is null; $ u \cdot u = 0$. 

The defect, $\mathbb{V}$, breaks the $\mathfrak{osp}(8^{\star}|4)$ algebra to $\mathfrak{osp}(4^{\star}|2) \oplus \mathfrak{osp}(4^{\star}|2)$. The coordinates are split into two parts; the parallel, $x^{a=1,2}$, and transverse, $x^{i=3,4,5,6}$. The original conformal symmetry $\mathfrak{so}(2,6)$ is broken to $\mathfrak{so}(2,2) \oplus \mathfrak{so}(4)$. The $\mathfrak{so}(2,2)$ part of the conformal sub-group preserved in the presence of $\mathbb{V}$ acts on the former set of coordinates, while the orthogonal rotations described by the $\mathfrak{so}(4)$ act on the latter. The presence of $\mathbb{V}$ results, also, in a breaking of the R-symmetry to $\mathfrak{so}(4)_{\text{R}}$. The embedding of $\mathfrak{so}(4)_{\text{R}}$ into the original $\mathfrak{so}(5)_{\text{R}}$ is via a polarization vector, $\theta$, such that $\theta^2 = 1$. 

We define the conformal cross-ratios via:
\begin{equation}\label{eq: cross_ratios}
\frac{x_{12}^2}{|x_1^i||x_2^i|}=\frac{(1-x)(1-\bar{x})}{\sqrt{x\bar{x}}}
\,          ,
\qquad 
\frac{x_1^jx_2^j}{|x_1^i||x_2^i|}=\frac{x+\bar{x}}{2\sqrt{x\bar{x}}}
\,          ,
\end{equation}
while the R-symmetry cross-ratio is given by:
\begin{equation}\label{eq: R_cross_ratio}
\sigma
= \frac{u_1 \cdot u_2}{(u_1 \cdot \theta)(u_2 \cdot \theta)}
= - \frac{(1-\omega)^2}{2\omega}
\,          .
\end{equation}

The conformal cross-ratios have a physical interpretation. We can use a conformal transformation to fix one point at $1$, while the second point is expressed in terms for the cross-ratios $x$ and $\bar{x}$. Then, the defect $\mathbb{V}$ intersects perpendicularly at $0$ and $\infty$. The $\{x,\bar{x}\}$ cross-ratios have the interpretation of the location of one of the operators \cite{Meneghelli:2022gps}. So, we have moved the two operators to lie on a $2$d plane. 
%\begin{equation}
%x_1 = (1,0,0,0,0,0)
%\,          ,
%\qquad
%x_2 = ((x+\bar{x})/2,-i(x-\bar{x})/2,0,0,0,0)
%\,          ,
%\end{equation}

In terms of these cross-ratios the defect $2$-point function can be expressed as \cite{Chen:2023yvw}:
\begin{align}\label{eq: defect_2point}
\llangle S_{k_1} S_{k_2} \rrangle
 = \frac{(u_1 \cdot \theta)^{k_1} (u_2 \cdot \theta)^{k_2}}{|x_1^i|^{2k_1} |x_2^i|^{2k_2}}
   \mathcal{F}(x,\bar{x},\sigma) 
\,          .
\end{align}

The correlator is a degree-$k_m$ polynomial in the $\sigma$-variable:
\begin{equation}
\mathcal{F}(x,\bar{x},\sigma) = \sum_{n=0}^{k_m} \sigma^n \mathcal{F}_n(x,\bar{x}) 
\,          ,
\end{equation}
where in the above the extremality is defined to be $k_m = \texttt{min}{(k_1,k_2)}$. Throughout this work, we will assume that $k_1 \geq k_2$ and consequently $k_m = k_2$ without loss of generality. 

Thus far, we have only utilized the bosonic part of the unbroken superconformal symmetry. A further condition is introduced by the fermionic generators, the superconformal Ward identity \cite{Meneghelli:2022gps}:
\begin{equation}\label{eq: ward_identities}
\left(\partial_x + \partial_\omega\right) 
\mathcal{F}(x,\bar{x},\omega)|_{x=\omega}
= 0 
\,              ,
\end{equation}
with its counterpart under $x \leftrightarrow \bar{x}$. A direct consequence of \cref{eq: ward_identities} is that upon performing the ``twist'' $\omega = \bar{x}$ we obtain:
\begin{equation}\label{eq: chiral_correlator}
\mathcal{F}(x,\bar{x},\bar{x}) = \zeta(x) 
\,          ,
\end{equation}
and likewise $\omega = x$ results in $\mathcal{F}(x,\bar{x}, x)=\zeta(\bar{x})$. 

The most general solution to the superconformal Ward identity is:
\begin{equation}\label{eq: sltn_ward_identity}
\mathcal{F}(x,\bar{x},\omega)=
\mathcal{F}_{\text{prot}}(x,\bar{x},\omega)
+
\frac{(x-\omega)(\bar{x}-\omega)(x-\omega^{-1})(\bar{x}-\omega^{-1})}{x \bar{x}} ~ \mathcal{H}(x,\bar{x},\omega)
\,          ,
\end{equation}
with $\mathcal{F}_{\text{prot}}(x,\bar{x},\omega)$ the protected part of the correlator:
\begin{equation}\label{eq: prot_corr}
\mathcal{F}_{\text{prot}}(x,\bar{x},\omega)
=  
\frac{(x-\omega)(x-\omega^{-1})}{(x-\bar{x})(x-\bar{x}^{-1})} \zeta(x) 
+ 
\frac{(\bar{x}-\omega)(\bar{x}-\omega^{-1})}{(\bar{x}-x)(\bar{x}-x^{-1})} \zeta(\bar{x}) 
\,          .
\end{equation}
The function $\mathcal{H}$ in \cref{eq: sltn_ward_identity} is the so-called reduced correlator; a degree-$(k_m-2)$ polynomial in $\sigma$ that can be found in \cite{Chen:2023yvw}, however, since it is not necessary for our considerations we omit its explicit expression. 

It is clear from \cref{eq: sltn_ward_identity} that upon performing the R-symmetry ``twist'', $\omega=\bar{x}$, we are left only with the protected part of the correlator, $\mathcal{F}_{\text{prot}}$. Then, from \cref{eq: prot_corr} we see that under the same ``twist'' we are left only with a holomorphic function, $\zeta(x)$. Based on the observation of \cite{Beem:2013sza,Beem:2014kka}, this holomorphic function is not an arbitrary function, but rather a $4$-point correlator of the underlying $2$d chiral algebra.

Before proceeding any further we want to present some basic facts about the chiral algebra conjecture and the $\mathcal{W}$-algebras that are pertinent to the $6$d, $\mathcal{N}=(2,0)$ theory. Since our presentation will be brief we refer the interested reader to appropriate sources for further details. For an excellent exposure to the chiral algebra conjecture we suggest the wonderful lectures \cite{Lemos:2020pqv}, and for a review and an introduction to $\mathcal{W}$-algebras we refer to \cite{watts1997,Bouwknegt:1992wg} respectively.

The authors of \cite{Beem:2014kka} demonstrated that there is an underlying description in terms of a $2$d chiral algebra for the $6$d, $\mathcal{N}=(2,0)$ theory. To obtain this description one has to pass to the cohomology of an appropriately chosen supercharge that is nilpotent of order $2$, $\mathbb{Q}^2 = 0$. This supercharge is given as a linear combination of the Poincar\'e and conformal supercharges. By passing to the cohomology what we mean is that we want to find operators, $\mathcal{O}$, that are $\mathbb{Q}$-closed, namely $\mathbb{Q}0=0$, but not $\mathbb{Q}$-exact, $\mathcal{O} \neq \mathbb{Q} Y$, see for example \cite{Lemos:2021azv}. Under this cohomological construction, denoted by $\chi$, the representations of the $\mathfrak{osp}(8^{\star}|4)$ algebra are mapped to $\mathfrak{sl}(2)$ representations. This has a manifestation at the level of the bulk-channel conformal blocks, as well, as they are re-organized into $\mathfrak{sl}(2)$ conformal blocks. 

In addition to the above, under the $\chi$-map the OPE coefficients of the $6$d theory reduce to the OPE coefficients of the operators that are associated with the appropriate representations occurring after the cohomological reduction. For the $6$d, $\mathcal{N}=(2,0)$ theories the chiral algebra is conjectured to be the well-known $\mathcal{W}$-algebra with the central charge being:
\begin{equation}\label{eq: c2d}
c_{2d} = 4N^3 - 3N - 1
\,          .
\end{equation}

The $\mathcal{W}$-algebra is generated by a set of primary operators, $W_p(z)$ with $p=2,3,\ldots, N$. These have a scaling weight $h=p$ and the lowest entry is identified with the holomorphic stress-energy tensor $T(z) \equiv W_2(z)$. Under the $\chi$-map, the $\tfrac{1}{2}$-BPS operators are matched to the primary operators that generate the algebra:
\begin{equation}
\chi: S_{k}(x,u) \mapsto W_k(z)
\,          .
\end{equation}  

The chiral algebra manages to, also, capture surface defect operators that are orthogonal to the chiral algebra plane. The reasoning behind this, is that the $\mathfrak{osp}(4^{\star}|2) \oplus \mathfrak{osp}(4^{\star}|2)$ algebra, which is the presented sub-algebra in the presence of $\mathbb{V}$, contains the supercharge $\mathbb{Q}$ that is used to define the cohomological reduction. The surface defect intersects the chiral algebra plane at two points, namely at $0$ and $\infty$. Under the $\chi$-map, $\mathbb{V}$ corresponds to the insertion of two vertex operators at those locations:
\begin{equation}
\chi: \mathbb{V} \mapsto \bar{V}(\infty)
\,          ,
\quad
\text{and}
\quad
V(0)
\,          .
\end{equation}  

This defines a module of the associated chiral algebra. The modules of the $\mathcal{W}$-algebras have been thoroughly studied in the past and in \cite{Meneghelli:2022gps} the authors showed that surface operators can be identified with a special class of modules, the so-called completely degenerate representations. Their structure can equivalently be translated in the OPE language \cite{Meneghelli:2022gps}:
\begin{equation}\label{eq: modules_OPEs}
\begin{aligned}
T(z) V(0) &\sim \frac{\Delta}{z^2} V(0) + \frac{2}{z} \partial V(0) 
\,          ,
\\
W_k(z) V(0) &\sim \frac{\omega_k}{z^k} + \frac{1}{z^{k-1}} \left( \frac{k ~ \omega_k}{2 \Delta} + V^{(k)}(0) \right) + \ldots
\,          ,
\end{aligned}
\end{equation}
where $\Delta$ in the above is given by the anomaly coefficient of the theory \cite{Meneghelli:2022gps} and $\omega_k$ is computed by a free-field realisation of the $\mathcal{W}$-algebra \cite{Fateev:1987zh}. 

In passing, we stress that there are two differences between the bulk and defect operators under the cohomological reduction. Defect operators are bound to the surface defect, $\mathbb{V}$, and therefore cannot be translated away from $0$ or $\infty$. Moreover, primary bulk operators are mapped to $\mathfrak{sl}(2)$ primaries, while primary defect operators can be mapped to descendants of the $\mathfrak{sl}(2)$-algebra, see \cite{Meneghelli:2022gps} for a more detailed explanation and explicit examples. 

All in all, we have argued that under the cohomological reduction, $\chi$, the $2$-point correlator of local $\tfrac{1}{2}$-BPS operators in the presence of a $\tfrac{1}{2}$-BPS defect, $\mathbb{V}$, considered in \cite{Chen:2023yvw} is translated to $4$-point correlator of two vertex operators and two primary generators of the chiral algebra:
\begin{equation}\label{eq: chiral_red_4points}
\chi: \llangle S_{k_1} S_{k_2} \rrangle \mapsto \langle\bar{V}(\infty) W_{k_1}(z) W_{k_2}(1) V(0) \rangle
\,              ,
\end{equation}
where in the above in addition to placing the two vertex operators at $0$ and $\infty$, since they cannot be moved away from these points, we have used conformal symmetry to move one of the local operators at $1$, while the other is left at $z$.
%%%%%%%%%%%%%%%%%%%%%%%%%%%%%%%%%%%%%%%%%%%%%%%%%%%%%%%%%%%%%%%%%%%%%%%%%%%%%%%%%%%%%%%%%%%%%%
\subsection{Review of results from supergravity}
The purpose of this section is to quickly review the main apsects of the SUGRA computation that was performed in \cite{Chen:2023yvw} in order to obtain all $2$-point correlation functions of two $\tfrac{1}{2}$-BPS operators in the presence of a $\tfrac{1}{2}$-BPS surface defect. 

The contribution to the connected defect $2$-point functions to leading order in the large-N limit is given as finite sum of Witten diagrams. This can be computed by following the standardized diagrammatic approach. A requirement of this approach is the precise form of the effective Lagrangian in AdS, and is an arduous task. One has to expand the defect effective AdS Lagrangian to quadratic order and extract the related Feynman rules. This has been explicitly done in closely related setups; in \cite{Gimenez-Grau:2023fcy} for the Wilson line, in \cite{Zhou:2024ekb} for the $\mathcal{N}=4$ on the real-projective space and in \cite{Chen:2025yxg} for giant gravitons. A common feature of these computations is that there are subtleties for the contact interactions. 

The above can be circumvented by making an ansatz in terms of all possible Witten diagrams whilst keeping the coefficients undetermined. In order to compute these coefficients we can use the power of superconformal symmetry. We include in the ansatz all possible diagrams, namely the bulk exchanges, defect exchanges and contact terms:
\begin{equation}
    \mathcal{F}_{\text{ansatz}}=    \mathcal{F}_{\text{exchange}}^{\text{bulk}}+\mathcal{F}_{\text{exchange}}^{\text{defect}}+\mathcal{F}_{\text{contact}}
\,          ,
\end{equation}
where in the above the contact term, $\mathcal{F}_{\text{contact}}$, includes contact Witten diagrams with zero derivatives only, since contact Witten diagrams with higher derivatives are more dominant than the exchange diagrams in the Regge limit \cite{Chen:2023yvw}. 

Importantly the set of fields that are allowed to be exchanged is finite and constrained. The conditions that are imposed in order to constrain the ansatz are the following \cite{Chen:2023yvw}: 
\begin{itemize}
    \item Selection rules following from the R-symmetry
    \item Non-extremal exchanges 
    \item The bulk-exchange Witten diagrams have even Lorentz spin
    \item The set of exchanged fields in the defect-channel contains only two fields
\end{itemize}

In addition to the above, owing to the spectrum of the theory the bulk and defect exchange Witten diagrams can be re-written as a finite sum of contact diagrams, that are known explicitly \cite{Rastelli:2017ecj,Gimenez-Grau:2023fcy}. 

It is now a straightforward exercise to impose the superconformal Ward identities given by \cref{eq: ward_identities} on the ansatz for any values of $k_1$ and $k_2$. The outcome of this computation is that all unknown coefficients of the ansatz are fixed up to an overall number in each case. An important observation comes from the fact that the parameters in the ansatz have the interpretation of OPE coefficients. Since the same coefficients appear in the study of different correlators, by considering $2$-point correlation functions of different $k_1$ and $k_2$, the overall constants that appear in each case can be reduced to a single one; the one of $k_1 = k_2 = 2$. This last coefficient can be determined in terms of the central charge of the theory, see e.g \cite{Meneghelli:2022gps}. Hence, this strategy fully fixes all $2$-point functions. In the result we can perform the twist $\omega = \bar{x}$ in order to obtain a meromorphic function given by \cite{Chen:2023yvw}:
\begin{equation}\label{eq: chiral_corrr}
\zeta(x) 
=  
\frac{1}{2} ~ b_{k_1\mathcal{D}} ~ b_{k_2\mathcal{D}}
\sum_{i=1}^{k_m-1} C_i ~ \left(- \frac{(x-1)^2}{x}\right)^{-i}
\,          ,
\end{equation}
with $C_i$ being the Catalan number expressed as $C_i = \tfrac{1}{i+1}\binom{2i}{i}$ and the defect OPE coefficients are given by \cite{Chen:2023yvw}:
\begin{equation}
b_{k \mathcal{D}} = \frac{k \Gamma(k)}{\sqrt{2^k \Gamma(2k-1)}}
\,          .
\end{equation}

The function $\zeta(x)$ has an interesting in-built symmetry. It is invariant under $x \rightarrow 1/x$. The physical interpretation of this crossing is that it corresponds to the exchange of the two bulk operators \cite{Meneghelli:2022gps} and this feature is pivotal in the forthcoming analysis. It is precisely this crossing symmetry that we will exploit in order to use only half of the OPE information; the bulk-channel OPE.

Instead of using the $\{x,\bar{x}\}$ cross-ratios there is a more natural choice that can be used given by \cite{Chen:2023yvw}:
\begin{equation}\label{eq: def_of_X}
X = -\frac{(x-1)^2}{2x}
\,          .
\end{equation}  
This makes crossing symmetry manifest, since under the $x \rightarrow \tfrac{1}{x}$ shift, $X$ remains invariant.
%%%%%%%%%%%%%%%%%%%%%%%%%%%%%%%%%%%%%%%%%%%%%%%%%%%%%%%%%%%%%%%%%%%%%%%%%%%%%%%%%%%%%%%%%%%%%%
\section{Bootstrapping the chiral correlator}\label{sec: bootstrap}
In this section we will see that we can obtain the result given by \cref{eq: chiral_corrr} by utilizing the underlying chiral algebra description of the theory, under which the defect $2$-point function is mapped to a $4$-point correlator. We can bootstrap the result using bulk-channel data only, which is a special feature of our setup, tied to the fact that the correlator has an in-built crossing symmetry which we exploit in order to use only half of the OPE information; the bulk-channel OPE. After obtaining the chiral correlator, we will perform a defect-channel expansion and obtain new predicitions, namely the defect-channel CFT data.  

We have already explained that the $2$-point correlator of local operators in the presence of the defect can be thought of as a $4$-point correlator in the chiral algebra
\begin{equation}\label{eq: def_chiral_fourpoint}
\langle\bar{V}(\infty) W_{k_1}(z) W_{k_2}(1) V(0) \rangle
\,          .
\end{equation}

Since we will be working with a $4$-point correlator, it is necessary to establish the relation between the defect $2$-point functions and the $4$-point functions in the chiral algebra.

When we performed the ``twist'' $\omega = \bar{x}$ to obtain the meromorphic function $\zeta(x)$ given by \cref{eq: chiral_correlator}, there is a kinematic factor that was implicitly extracted. However, it is not clear what this factor is and it becomes important, since the $\mathfrak{sl}(2)$-blocks depend on this choice.  

Let us consider the holomorphic part of a generic $4$-point function of scalar operators in $2$d, that have dimensions $h_i$. The form of the $4$-point function is dictated by conformal covariance:
\begin{equation}\label{eq: def_G}
G(x_i) = 
\frac{1}{\left(x^2_{12}\right)^{\tfrac{h_1+h_2}{2}}\left(x^2_{34}\right)^{\tfrac{h_3+h_4}{2}}}
\left(\frac{x^2_{14}}{x^2_{24}}\right)^{\tfrac{h_1 - h_2}{2}}
\left(\frac{x^2_{14}}{x^2_{13}}\right)^{\tfrac{h_3 - h_4}{2}}\mathcal{G}(z)
\,          ,
\end{equation}
where in the above $z$ is the conformal cross-ratio given by:
\begin{equation}\label{eq: def_cross_z}
z = \frac{x_{12}x_{34}}{x_{13}x_{24}}  
\,      .
\end{equation}

The conformal block decomposition of the correlator reads: 
\begin{equation}\label{eq: sl2_decompose}
\mathcal{G}(z) = \sum_{i} C_{h_1 h_2 h_i} C_{h_3 h_4 h_i} ~ h^{(h_{12},h_{34})}_{h_i}(z)
\,          ,
\end{equation}
with the $\mathfrak{sl}(2)$-blocks being given by:
\begin{equation}
h^{(h_{12},h_{34})}_{h_i}(z) = z^{h_i} ~ {}_2F_{1}\left(h_i - h_{12}, h_i + h_{34}, 2h_i  ; z\right)
\,      ,
\end{equation}
and we have used the abbreviation $h_{ij} = h_i - h_j$.

The twisted correlator is obtained from the defect $2$-point function. Hence, it should be the ratio of $G$ defined above and the $2$-point function of vertex operators. If we place the vertex operators at the positions $x_3$ and $x_4$, we should be focusing on the combination: 
\begin{equation}
\frac{G(x_i)}{\left(x^2_{34}\right)^{-h_3}} 
\,          .
\end{equation}

Setting $h_3=h_4$ in the definition, \cref{eq: def_G}, we get: 
\begin{equation}
\frac{G}{\left(x^2_{34}\right)^{-h_3}} =
\frac{1}{\left(x^2_{12}\right)^{\tfrac{h_1+h_2}{2}}}
\left(\frac{x^2_{14}}{x^2_{24}}\right)^{\tfrac{h_1 - h_2}{2}}
\mathcal{G}(z)
\,          .
\end{equation}
We can use conformal symmetry to fix $x_1 = 1, x_3 = 0, x_4 = \infty$, while $x_2 = x$ such that they correspond to the chiral-algebra plane. The $4$-point function is now: 
\begin{equation}\label{eq: 4_point_gauge}
\frac{G}{\left(x^2_{34}\right)^{-h_3}} =
\frac{1}{(1-x)^{\tfrac{h_1+h_2}{2}}}
\mathcal{G}(1-x)
\,          .
\end{equation}

Let us go back to the $2$-point correlator. In the defect-free case, the $2$-point function of local operators is given by: 
\begin{equation}
\langle S_k(x_1, u_1) S_k(x_2, u_2) \rangle  = \frac{(u_1 \cdot u_2)^k}{(x_{12}^2)^{2k}}
\,          ,
\end{equation}
and we know that under the cohomological reduction this becomes 
\begin{equation}
\chi: \langle S_k(x_1, u_1) S_k(x_2, u_2) \rangle \rightarrow \langle W_k(z_1) W_k(z_2) \rangle = \frac{1}{z^{2k}_{12}}
\,          .
\end{equation}
From the above we can determine how the ``twist'' acts on the combination $u_1 \cdot u_2$
\begin{equation}\label{eq: action_u1u2}
\chi: u_1 \cdot u_2 \rightarrow \bar{z}^2_{12}
\,          ,
\end{equation}
having used that in $2$d, we have $x^2_{12} = z^2_{12} \bar{z}^2_{12}$. 

Let us now determine how the ``twist'' acts on the combination $(u_a\cdot \theta)/|x_a^i|^2$. Starting from the combination: 
\begin{equation}
\frac{(u_1 \cdot \theta)(u_2 \cdot \theta)}{|x_1^i|^2|x_2^i|^2} = -\frac{2\omega }{(1-\omega)^2}\frac{(1-x)^2(1-\bar{x})^2}{x\bar{x}}\frac{u_1\cdot u_2}{x_{12}^4}
\rightarrow - 2 \frac{(1-x)^2}{x} \frac{1}{z^2_{12}} = - \frac{2}{x}
\,          ,
\end{equation}
where in the first step we have inserted the resolution of the identity twice, subsequently used the definitions for the conformal and R-symmetry cross ratios, see \cref{eq: cross_ratios,eq: R_cross_ratio}, then performed $\omega = \bar{x}$ ``twist'' and used \cref{eq: action_u1u2}, and finally we gauge fixed the two points at $1$ and $x$. From the above we conclude: 
\begin{equation}\label{eq: u_theta_x}
\frac{u_a\cdot \theta}{|x_a^i|^2}\to \frac{(-2)^{\frac{1}{2}}}{z_i}
\,          .
\end{equation}
We note that there is a more direct way to obtain the above result that we do not pursue here. It consists of finding explicitly the dependence of the polarisation vectors $u$ in terms of the cross-ratios $\{z,\bar{z}\}$ and then obtain the action of the R-symmetry ``twist''. 

Therefore, the defect $2$-point function under the ``twist'' becomes: 
\begin{equation}\label{eq: twisted_2_point}
\chi: \llangle S_{k_1} S_{k_2} \rrangle \rightarrow \frac{(-2)^{\tfrac{k_1 + k_2}{2}}}{x^{k_2}} \zeta(x)
\,          ,
\end{equation}
where we have used \cref{eq: u_theta_x} and we have fixed one point at $1$ and the other at $x$. 

Requiring that \cref{eq: twisted_2_point} is equal to \cref{eq: 4_point_gauge} we obtain: 
\begin{equation} 
\mathcal{G}(z)=(-2)^{\frac{k_1+k_2}{2}}\frac{(1-x)^{k_1+k_2}}{x^{k_2}}\zeta(x)
\,          ,
\end{equation}
with $x=1-z$\footnote{This can be seen by using conformal symmetry to fix $x_1 =1, x_3 = 0, x_4 = \infty$, while leaving $x_2 =x$ in the definition of cross-ratio, see \cref{eq: def_cross_z}.}. This, also, fixes the issue of normalisation. To pass from the ``un-twisted'' OPE coefficients to the ``twisted'' ones, $3$-point couplings reduce with unit coefficient, while defect $1$-point functions pick up an extra factor of $(-2)^{\tfrac{k}{2}}$. 

Combining the above results we obtain the following expression for the $2$d decomposition in terms of $6$d CFT data: 
\begin{equation}\label{eq: 2d_6d}
\frac{(1-x)^{k_1 + k_2}}{x^{k_2}} \zeta(x) = \sum_{k} (-2)^{\tfrac{k-k_1-k_2}{2}} \lambda_{k_1 k_2 k} a_k ~ h^{(k_{12},0)}_{k}(z)
\,          .
\end{equation}

In passing we mention that there is another way of obtaining the factor of $(-2)^{\tfrac{k-k_1-k_2}{2}}$ that appears in \cref{eq: 2d_6d}. Following \cite{Chen:2023yvw} we can construct the superconformal blocks of short multiplets. In these conventions the normalisation is such that the conformal block of the exchanged field has unit coefficient. We, then, single out the single-trace part of the blocks and perform the R-symmetry ``twist'', $\omega = \bar{x}$, after which we end up obtaining the standard $\mathfrak{sl}(2)$-blocks multiplied by the $(-2)^{\tfrac{k-k_1-k_2}{2}}$ factor.
%%%%%%%%%%%%%%%%%%%%%%%%%%%%%%%%%%%%%%%%%%%%%%%%%%%%%%%%%%%%%%%%%%%%%%%%%%%%%%%%%%%%%%%%%%%%%%
\subsection{Ansatz and crossing symmetry}
Let us start by making an ansatz for the correlation function. From the perspective of the chiral algebra, this is a $4$-point correlator that is a meromorphic function. The poles of the function are at such locations that they appear as $z$ approaches the other points. The order of the poles and the residues of the function are dictated by the OPE. This basic idea, that the meromorphic function can be determined by the singularities that are controlled by the OPE of the chiral algebra has been used previously \cite{Headrick:2015gba,Rastelli:2017ymc,Behan:2021pzk,Meneghelli:2022gps}. 

Our ansatz is given by the following:
\begin{equation}\label{eq: ansatz}
f(z) 
=
\sum^{2k_m - 2}_{j=0} \alpha_j \frac{(1-z)^j}{z^{2k_m - 2}}
\,          ,
\end{equation}
and we impose invariance under $z \rightarrow - z/(1-z)$\footnote{This is just the $x \rightarrow 1/x$ symmetry of the $2$-point function expressed in the language of the chiral $4$-point correlator.}. We solve this condition and the solution to this is given by:
\begin{equation}\label{eq: cross_sltn}
\alpha_{2k_m-2-r} = \alpha_r
\,          ,
\quad
\text{with}
\quad
r=0,1,2,\ldots,k_m-2
\,          .
\end{equation}  

However, there is a more natural way to write the ansatz, namely in terms of the $Z$ cross-ratio:
\begin{equation}\label{eq: crossing_ansatz_2}
g(z) = \sum^{k_m - 1}_{i=0} \beta_i Z^{-i}
\,          ,
\end{equation}
with $Z$ being given by:
\begin{equation}
Z = - \frac{z^2}{2(1-z)}
\,          ,
\end{equation}  
which is the analogous cross-ratio of $X$ defined in \cref{eq: def_of_X}. 

The question that arises is if the two ansatze for the correlators, given by \cref{eq: ansatz,eq: crossing_ansatz_2}, are equivalent. In other words, we have to determine if we can always solve the coefificients $\alpha_j$ that appear in \cref{eq: ansatz} in such a way that we obtain \cref{eq: crossing_ansatz_2}, subject to the condition for crossing invariance. In order to address the above question we need to consider the ansatz given by \cref{eq: ansatz} with the solution obtained in \cref{eq: cross_sltn} imposed on its various coefficients and set it to be equal to \cref{eq: crossing_ansatz_2}. The task at hand is to solve for all $\beta_i$. Then, try to obtain an expression for the numerical coefficients of each one of the $\alpha_i$-coefficients. 

Let's consider a concrete example in order to exemplify our logic. We make the choice that extremality is given by $k_m=13$. Following the logic we outlined above, we determine the $\alpha_0$ to be given by: 
\begin{equation}
\alpha_0 = 
-
\frac{3 \cdot 2^{i-21}}{\sqrt{\pi}}
(i-23)_{11} 
\frac{\Gamma \! \left( i - \frac{23}{2} \right)}{\Gamma \! \left(i+1\right)} 
\beta_0
\,          ,
\end{equation}
where in the above $(a)_b$ is the Pochhammer symbol. 

We can work in the same manner that we did for $\alpha_0$ for $\alpha_1$, and subsequently the other coefficients trying to spot a pattern. As it turns out it is sufficient to obtain the expressions for $\alpha_0, \alpha_1, \ldots, \alpha_6$ in order to conjecture the general formula -general in terms of the $j$-index since extremality is still fixed to the value $k_m = 13$- which is given by: 
\begin{equation}
\alpha_j = 
\left[
\frac{(-1)^{j} ~ 2^{-23+i}}{\sqrt{\pi}}~ (j-12) ~ \frac{\Gamma \! \left(i-12\right)\Gamma \! \left(i-\tfrac{23}{2}\right)}{\Gamma \! \left(i-j + 1\right)\Gamma \! \left(i+j-23\right)}
+(-2)^{-i} \delta_{i,j,12}
\right]
\beta_i
\,          ,
\end{equation}
where in the above $\{i,j\}=0,1,2,\dots,k_m-1$. 

Of course, having worked out one concrete case of extremality is not enough to be able to guide us in order to write the general expression. However, using the cases $k_m = \{13,14,15\}$ we were able to deduce:
\begin{equation}\label{eq: generalb_correct}
\begin{aligned}
\alpha_j = 
&\left[
(-1)^j ~ 2^{2-i} ~ (j-(k_m-1)) ~ 
\frac{\Gamma \! \left(2i-(2k_m-2) \right)}{\Gamma \! \left(i-j+1\right) \Gamma \! \left(i + j -(2k_m-3)\right)} 
+ \right. \\
&\quad \left. {}
(-2)^{-i} \delta_{i,j,k_m-1}
\vphantom{\frac{1}{2}}
\right]
\beta_i
\,          .
\end{aligned}
\end{equation}

Hence, we have concluded that the ansatze given by \cref{eq: ansatz,eq: crossing_ansatz_2} are equivalent with their coefficients being related via \cref{eq: generalb_correct}. Therefore, we can always re-write the chiral correlator in the form given by \cref{eq: crossing_ansatz_2}. 
%%%%%%%%%%%%%%%%%%%%%%%%%%%%%%%%%%%%%%%%%%%%%%%%%%%%%%%%%%%%%%%%%%%%%%%%%%%%%%%%%%%%%%%%%%%%%%
\subsection{Fixing the ansatz using only bulk-channel data}
We are at a position to fully fix the ansatz given by \cref{eq: crossing_ansatz_2}. To do so, we need to perform the bulk-channel expansion, $z \rightarrow 0$. The decomposition of the ansatz is:
\begin{equation}\label{eq: bulk_decompose}
\frac{z^{k_1 + k_2}}{(1-z)^{k_2}} \sum^{k_m - 1}_{i=0} \beta_i Z^{-i}
= 
\sum_{k=k_{\texttt{min}}}^{k_{\texttt{max}}} (-2)^{\tfrac{k-k_1-k_2}{2}} a_k ~ \lambda_{k_1 k_2 k} ~ h^{(k_{12},0)}_k(z)
\,          ,
\end{equation}
where we have used the relation of cross-ratios, $x=1-z$, in \cref{eq: 2d_6d} in order to obtain the correct pre-factor. In the above the sum starts from $k_{\texttt{min}} = |k_{12}|+2$ and goes to $k_{\texttt{max}}=k_1 + k_2 -2 $ in steps of $2$.

The defect $1$-point couplings are given by \cite{Corrado:1999pi}:
\begin{equation}
a_k = \frac{\Gamma (k)}{\sqrt{2^{k} \Gamma (2 k-1)}} 
\,          ,
\end{equation}
while the $3$-point couplings of the theory are: \cite{Corrado:1999pi,Bastianelli:1999en}:
\begin{equation}
\lambda_{k_1,k_2,k} = \frac{2^{k_1 + k_2 + k - 2}
        \Gamma \! \left(\frac{k_1 + k_2 + k}{2}\right)}
       {\pi^{3/2}} 
\frac{\Gamma \! \left(\frac{-k_1 + k_2 + k + 1}{2}\right) \Gamma \! \left(\frac{k_1 - k_2 + k + 1}{2}\right) \Gamma \! \left(\frac{k_1 + k_2 - k + 1}{2}\right)}{\sqrt{\Gamma \! \left(2 k_1-1\right) \Gamma \! \left(2 k_2-1\right) \Gamma \! \left(2 k-1\right)}}
\,          .
\end{equation}

Let us start by considering a concrete example, in order to work out the details of the above. We focus on the simplest case, namely $k_1 = k_2 =2$. Expanding both sides of \cref{eq: bulk_decompose} we obtain: 
\begin{equation}
-2 \beta_1 z^2 = - \frac{z^2}{4} 
\,          ,
\end{equation}
from which we can easily deduce that $\beta_1 = 1/8$. Hence, we have obtained that in this case the chiral correlator is: 
\begin{equation}
g(z) = - \frac{1-z}{4z^2}
\,          .
\end{equation}
After shifting the $4$-point to the $2$-point cross-ratio, using $z=1-x$, the above solution agrees with \cref{eq: chiral_corrr} as it should. We proceed to work out more cases until we were able to spot a pattern. We find: 
\begin{equation}\label{eq: bootstrap_sltn}
g(z) 
=  
\frac{1}{2} ~ b_{k_1\mathcal{D}} ~ b_{k_2\mathcal{D}}
\sum_{i=1}^{k_m-1} \frac{1}{i+1} ~ \binom{2i}{i} ~ (2 Z)^{-i}
\,          .
\end{equation}
In the above, however, $\tfrac{1}{i+1}\binom{2i}{i}$ is just the Catalan number, and hence the solution to our bootstrap program agrees perfectly with \cref{eq: chiral_corrr}. We end this section by mentioning that we have cross-checked \cref{eq: bootstrap_sltn} against many cases that were not used in order to obtain this formula.

At this point we wish to consider another approach of fixing the chiral correlator. This was briefly mentioned in \cite{Chen:2023yvw}. As was pointed out by the authors of that paper, there is another efficient way to fix the ansatz. We choose to work with the $\{x,\bar{x}\}$ cross-ratios here since our comment will be brief. The logic is to use the defect-channel expansion, $x \rightarrow 0$, of the ansatz while requiring that the leading term of the expansion $x$ is followed only by $x^{k_m}$ and higher-order ones. The absence of the terms between the first, $x$, and the next one, $x^{k_m}$, is due to the fact that the exchange Witten diagrams in the defect channel have only one supermultiplet of fields and the spectrum of protected operators has a gap. As we shall see, this is enough to fix the ansatz up to a number.

Let us consider an example to demonstrate how this process works. Let us consider a $2$-point function of extremality $k_m = 4$. The ansatz is given by \cref{eq: crossing_ansatz_2}, and we use $z=1-x$. This results in the ansatz being written with $Z$ replaced by $X$, with the latter being defined in \cref{eq: def_of_X}. For $k_m = 4$ this is explicitly written out as: 
\begin{equation}
\beta_0 - 2 \beta_1 \frac{x}{(1-x)^2} + 4 \beta_2  \frac{x^2}{(1-x)^4} - 8 \beta_3 \frac{x^3}{(1-x)^6}
\,          .
\end{equation}
We perform the $x \rightarrow 0$ expansion of the above and we impose the conditions that the leading term is the one scaling like $x$ and that there are no $x^2$ and $x^3$ terms in the expansion. This gives us the solution 
\begin{equation}
\beta_0 = 0
\,      ,
\qquad
\beta_2 = \beta_1
\,      ,
\qquad
\beta_3 = \frac{5}{4} \beta_1
\,          .
\end{equation}
We can repeat the process above for different values of $k_m$ and try to spot a pattern for the overall number that relates the various coefficients to $\beta_1$. This is not too difficult and a few choices of $k_m$ are sufficient. The solution we obtain is given by:
\begin{equation}
2 \beta_1 \sum^{k_m - 1}_{i=1} C_i \left(2 X\right)^{-i}
\,          ,
\end{equation}
where in the above $C_i$ is just the Catalan number. Having fixed the correlator up to a number the final question that remains is how to fix the number. This can be fixed by identifying this number with the defect-channel data obtained in \cite{Chen:2023yvw} and the large-N expansion of \cref{eq: c2d}. This results in $\beta_1 = \tfrac{1}{4} ~ b_{k_1\mathcal{D}} ~ b_{k_2\mathcal{D}}$.

Before proceeding to extract the dCFT data we wish to clarify an important point. The fact that there are missing powers in the defect-channel expansion, resulting from the spectrum of this particular defect set-up, is not important for the strategy we employed in order to bootstrap the correlator using the bulk-channel data. More specifically, the approach of \cite{Meneghelli:2022gps,Rastelli:2017ymc,Behan:2021pzk} that we generalised here by including a defect and higher KK-modes for the local operators can be applied in similar set-ups with different spectra that perhaps do not have missing powers in the defect-channel expansion.   
%%%%%%%%%%%%%%%%%%%%%%%%%%%%%%%%%%%%%%%%%%%%%%%%%%%%%%%%%%%%%%%%%%%%%%%%%%%%%%%%%%%%%%%%%%%%%%
\subsection{CFT data in the defect channel}
In this section we will extract the OPE data related to the defect operator expansion. To do so, we need to expand the chiral correlator, \cref{eq: chiral_corrr}, in defect-channel conformal blocks. These blocks are simply given by the monomials in $x$. Hence, we consider the decomposition of $\zeta(x)$:
\begin{equation}
\zeta(x) = \sum^{\infty}_{n=0} \gamma_n x^n
\,              ,
\end{equation}   
and perform the defect-channel expansion, $x \rightarrow 0$. Solving for the $\beta_n$-coefficients yields: 
\begin{equation}\label{eq: dOPE_data}
\gamma_n = - \frac{(-1)^{k_2}  b_{k_1\mathcal{D}} b_{k_2\mathcal{D}}}{k_2 ! (k_2-2)!} ~ n ~  (n-k_2+1)_{k_2-2} ~ (n+2)_{k_2+2}
\,          .
\end{equation}
For the special case given by $k_1 = 2$ and $k_2 = 2$, \cref{eq: dOPE_data} reduces to $-\tfrac{n}{4}$, which agrees with the result obtained in \cite{Meneghelli:2022gps} as it should. 

Finally, we note that \cref{eq: dOPE_data} is a new prediction that we obtain here. The leading term that scales like $x$, corresponds to single-trace operators. All the sub-leading ones, starting from $x^{k_m}$ and going higher to higher-scaling powers, represent double-trace contributions. 
%%%%%%%%%%%%%%%%%%%%%%%%%%%%%%%%%%%%%%%%%%%%%%%%%%%%%%%%%%%%%%%%%%%%%%%%%%%%%%%%%%%%%%%%%%%%%%
\section{Epilogue}\label{sec: final}
In this work we revisited the computation of the $2$-point correlation functions of $\tfrac{1}{2}$-BPS operators in the presence of a $\tfrac{1}{2}$-BPS defect in the $6$d, $\mathcal{N}=(2,0)$ theory. We have seen how to use the underlying chiral algebra in order to bootstrap the chiral correlator, dubbed $\zeta(x)$, using bulk-channel data only. We proceeded to use the defect-channel expansion of the correlator in order to provide a wealth of new CFT data in the defect channel. 

Our computation can be seen as a natural continuation and extension of \cite{Meneghelli:2022gps} where the authors considered the special case $k_1 = k_2 = 2$ and, also, serves as a concrete and non-trivial consistency check of the supergravity computation performed in \cite{Chen:2023yvw}. 

Another importance of understanding in detail how the chiral algebra works particualrly in this set-up is that it provides us with exact results despite the fact that the theory is non-Lagrangian. This is in sharp contrast with Lagrangian theories where supersymmetric localisation can be applied. Therefore, employing the chiral algrebra can, also, be seen as a viable alternative to supersymmetric localisation for non-Lagrangian theories. 

It would be interesting to extend the results of this work and also apply the ideas developed here in related set-ups. We mention some exciting future avenues below. 

A fascinating arena where similar considerations can be explored is given by the $4$-point correlation functions of two maximal giant gravitons and two light $\tfrac{1}{2}$-BPS operators in the context of the $\mathcal{N}=4$ super Yang-Mills that were obtained recently in \cite{Chen:2025yxg}. This set-up is particularly enticing as it is the first one where a hidden conformal symmetry of the defect correlators has been explicitly realized. 

Forbye, there are other surface defects that one can consider in the $6$d, $\mathcal{N}=(2,0)$ theory that span an $\text{AdS}_3 \times \text{S}^3$ subspace in the original theory. This $\text{AdS}_3 \times \text{S}^3$ sub-manifold can either be fully inside $\text{AdS}_7$ or the $\text{S}^3$ can be a sub-space of the original $S^4$. These surface defects are described by probe $\text{M}5$-branes, see e.g \cite{Beccaria:2024gkq}. It would be interesting to extend the approach that we have employed here as well as the approach of \cite{Chen:2023yvw} to study the $2$-point functions in that case. 

Returning to the set-up that was the focus of this work, of course, it would be highly interesting to utilize a combination of the light-cone OPE and the chiral algebra in order to obtain higher-point functions in the same vein as \cite{Goncalves:2025jcg}. 

Finally, an important direction that we wish to mention is the use of the flat-space limit for defect correlators that was developed in \cite{Alday:2024srr} in order to obtain information beyond the supergravity regime, probing M-theory. 
%%%%%%%%%%%%%%%%%%%%%%%%%%%%%%%%%%%%%%%%%%%%%%%%%%%%%%%%%%%%%%%%%%%%%%%%%%%%%%%%%%%%%%%%%%%%%%
\newpage
%%%%%%%%%%%%%%%%%%%%%%%%%%%%%%%%%%%%%%%%%%%%%%%%%%%%%%%%%%%%%%%%%%%%%%%%%%%%%%%%%%%%%%%%%%%%%%
\section*{Acknowledgments} 
We are grateful to Xinan Zhou for initial collaboration, a large-N number of discussions throughout the various stages of the project, and for reading a draft of this work and offering valuable comments and feedback and to Junding Chen for a discussion on superconformal blocks. We appreciate the warm hospitality at the Julius-Maximilians-Universit\"at W\"urzburg where parts of this work were performed. The work of KCR is supported by the Juan de la Cierva program through the fellowship JDC2023-052686-I, by the Xunta de Galicia (CIGUS Network of Research Centres and grant ED431C-2021/14), the European Union, the
Mar\'ia de Maeztu grant CEX2023-001318-M funded by MICIU/AEI/10.13039/501100011033
and the Spanish Research State Agency (grant PID2023-152148NB-I00).
%%%%%%%%%%%%%%%%%%%%%%%%%%%%%%%%%%%%%%%%%%%%%%%%%%%%%%%%%%%%%%%%%%%%%%%%%%%%%%%%%%%%%%%%%%%%%%
\newpage
%%%%%%%%%%%%%%%%%%%%%%%%%%%%%%%%%%%%%%%%%%%%%%%%%%%%%%%%%%%%%%%%%%%%%%%%%%%%%%%%%%%%%%%%%%%%%%
\bibliographystyle{ssg}
\bibliography{draft}

\begingroup\raggedright\begin{thebibliography}{10}

\bibitem{Liendo:2012hy}
P.~Liendo, L.~Rastelli, and B.~C. van Rees, ``{The Bootstrap Program for
  Boundary CFT$_d$},'' {\em JHEP} {\bf 07} (2013) 113,
  \href{https://arxiv.org/abs/1210.4258}{{\tt 1210.4258}}.

\bibitem{Billo:2016cpy}
M.~Bill\`o, V.~Gon\c{c}alves, E.~Lauria, and M.~Meineri, ``{Defects in
  conformal field theory},'' {\em JHEP} {\bf 04} (2016) 091,
  \href{https://arxiv.org/abs/1601.02883}{{\tt 1601.02883}}.

\bibitem{Lemos:2017vnx}
M.~Lemos, P.~Liendo, M.~Meineri, and S.~Sarkar, ``{Universality at large
  transverse spin in defect CFT},'' {\em JHEP} {\bf 09} (2018) 091,
  \href{https://arxiv.org/abs/1712.08185}{{\tt 1712.08185}}.

\bibitem{Liendo:2019jpu}
P.~Liendo, Y.~Linke, and V.~Schomerus, ``{A Lorentzian inversion formula for
  defect CFT},'' {\em JHEP} {\bf 08} (2020) 163,
  \href{https://arxiv.org/abs/1903.05222}{{\tt 1903.05222}}.

\bibitem{Bissi:2018mcq}
A.~Bissi, T.~Hansen, and A.~S\"oderberg, ``{Analytic Bootstrap for Boundary
  CFT},'' {\em JHEP} {\bf 01} (2019) 010,
  \href{https://arxiv.org/abs/1808.08155}{{\tt 1808.08155}}.

\bibitem{Mazac:2018biw}
D.~Maz\'a\v{c}, L.~Rastelli, and X.~Zhou, ``{An analytic approach to
  BCFT$_{d}$},'' {\em JHEP} {\bf 12} (2019) 004,
  \href{https://arxiv.org/abs/1812.09314}{{\tt 1812.09314}}.

\bibitem{Barrat:2022psm}
J.~Barrat, A.~Gimenez-Grau, and P.~Liendo, ``{A dispersion relation for defect
  CFT},'' {\em JHEP} {\bf 02} (2023) 255,
  \href{https://arxiv.org/abs/2205.09765}{{\tt 2205.09765}}.

\bibitem{Bianchi:2022ppi}
L.~Bianchi and D.~Bonomi, ``{Conformal dispersion relations for defects and
  boundaries},'' {\em SciPost Phys.} {\bf 15} (2023), no.~2 055,
  \href{https://arxiv.org/abs/2205.09775}{{\tt 2205.09775}}.

\bibitem{Cuomo:2021kfm}
G.~Cuomo, Z.~Komargodski, and M.~Mezei, ``{Localized magnetic field in the O(N)
  model},'' {\em JHEP} {\bf 02} (2022) 134,
  \href{https://arxiv.org/abs/2112.10634}{{\tt 2112.10634}}.

\bibitem{Cuomo:2022xgw}
G.~Cuomo, Z.~Komargodski, M.~Mezei, and A.~Raviv-Moshe, ``{Spin impurities,
  Wilson lines and semiclassics},'' {\em JHEP} {\bf 06} (2022) 112,
  \href{https://arxiv.org/abs/2202.00040}{{\tt 2202.00040}}.

\bibitem{Giombi:2021uae}
S.~Giombi, E.~Helfenberger, Z.~Ji, and H.~Khanchandani, ``{Monodromy defects
  from hyperbolic space},'' {\em JHEP} {\bf 02} (2022) 041,
  \href{https://arxiv.org/abs/2102.11815}{{\tt 2102.11815}}.

\bibitem{Giombi:2020rmc}
S.~Giombi and H.~Khanchandani, ``{CFT in AdS and boundary RG flows},'' {\em
  JHEP} {\bf 11} (2020) 118, \href{https://arxiv.org/abs/2007.04955}{{\tt
  2007.04955}}.

\bibitem{Bianchi:2022sbz}
L.~Bianchi, D.~Bonomi, and E.~de~Sabbata, ``{Analytic bootstrap for the
  localized magnetic field},'' {\em JHEP} {\bf 04} (2023) 069,
  \href{https://arxiv.org/abs/2212.02524}{{\tt 2212.02524}}.

\bibitem{Gimenez-Grau:2022ebb}
A.~Gimenez-Grau, ``{Probing magnetic line defects with two-point functions},''
  \href{https://arxiv.org/abs/2212.02520}{{\tt 2212.02520}}.

\bibitem{Raviv-Moshe:2023yvq}
A.~Raviv-Moshe and S.~Zhong, ``{Phases of surface defects in Scalar Field
  Theories},'' {\em JHEP} {\bf 08} (2023) 143,
  \href{https://arxiv.org/abs/2305.11370}{{\tt 2305.11370}}.

\bibitem{SoderbergRousu:2023pbe}
A.~S\"oderberg~Rousu, ``{The O(N)-flavoured replica twist defect},'' {\em JHEP}
  {\bf 07} (2023) 022, \href{https://arxiv.org/abs/2304.08116}{{\tt
  2304.08116}}.

\bibitem{SoderbergRousu:2023nvd}
A.~S\"oderberg~Rousu, ``{The discontinuity method in a BCFT},''
  \href{https://arxiv.org/abs/2304.02271}{{\tt 2304.02271}}.

\bibitem{Trepanier:2023tvb}
M.~Tr\'epanier, ``{Surface defects in the O(N) model},'' {\em JHEP} {\bf 09}
  (2023) 074, \href{https://arxiv.org/abs/2305.10486}{{\tt 2305.10486}}.

\bibitem{Giombi:2023dqs}
S.~Giombi and B.~Liu, ``{Notes on a surface defect in the O(N) model},'' {\em
  JHEP} {\bf 12} (2023) 004, \href{https://arxiv.org/abs/2305.11402}{{\tt
  2305.11402}}.

\bibitem{Baerman:2024tql}
J.~Baerman, A.~Chalabi, and C.~Kristjansen, ``{Superconformal two-point
  functions of the Nahm pole defect in N=4 super-Yang-Mills theory},'' {\em
  Phys. Lett. B} {\bf 853} (2024) 138642,
  \href{https://arxiv.org/abs/2401.13749}{{\tt 2401.13749}}.

\bibitem{Chalabi:2025nbg}
A.~Chalabi, C.~Kristjansen, and C.~Su, ``{Integrable corners in the space of
  Gukov-Witten surface defects},'' {\em Phys. Lett. B} {\bf 866} (2025) 139512,
  \href{https://arxiv.org/abs/2503.22598}{{\tt 2503.22598}}.

\bibitem{deLeeuw:2023wjq}
M.~de~Leeuw, C.~Kristjansen, G.~Linardopoulos, and M.~Volk, ``{B-type anomaly
  coefficients for the D3-D5 domain wall},'' {\em Phys. Lett. B} {\bf 846}
  (2023) 138235, \href{https://arxiv.org/abs/2307.10946}{{\tt 2307.10946}}.

\bibitem{Georgiou:2023yak}
G.~Georgiou, G.~Linardopoulos, and D.~Zoakos, ``{Holographic correlators of
  semiclassical states in defect CFTs},'' {\em Phys. Rev. D} {\bf 108} (2023),
  no.~4 046016, \href{https://arxiv.org/abs/2304.10434}{{\tt 2304.10434}}.

\bibitem{Linardopoulos:2025ypq}
G.~Linardopoulos, ``{String theory methods for defect CFTs},''
  \href{https://arxiv.org/abs/2501.11985}{{\tt 2501.11985}}.

\bibitem{Linardopoulos:2025mav}
G.~Linardopoulos, ``{B-type anomaly coefficients for the D3-D7 domain wall},''
  \href{https://arxiv.org/abs/2502.13613}{{\tt 2502.13613}}.

\bibitem{Chiodaroli:2016jod}
M.~Chiodaroli, J.~Estes, and Y.~Korovin, ``{Holographic two-point functions for
  Janus interfaces in the $D1/D5$ CFT},'' {\em JHEP} {\bf 04} (2017) 145,
  \href{https://arxiv.org/abs/1612.08916}{{\tt 1612.08916}}.

\bibitem{Giombi:2017cqn}
S.~Giombi, R.~Roiban, and A.~A. Tseytlin, ``{Half-BPS Wilson loop and
  AdS$_2$/CFT$_1$},'' {\em Nucl. Phys. B} {\bf 922} (2017) 499--527,
  \href{https://arxiv.org/abs/1706.00756}{{\tt 1706.00756}}.

\bibitem{Wang:2020seq}
Y.~Wang, ``{Taming defects in $ \mathcal{N} $ = 4 super-Yang-Mills},'' {\em
  JHEP} {\bf 08} (2020), no.~08 021,
  \href{https://arxiv.org/abs/2003.11016}{{\tt 2003.11016}}.

\bibitem{Drukker:2020swu}
N.~Drukker, S.~Giombi, A.~A. Tseytlin, and X.~Zhou, ``{Defect CFT in the 6d
  (2,0) theory from M2 brane dynamics in AdS$_7 \times$S$^4$},'' {\em JHEP}
  {\bf 07} (2020) 101, \href{https://arxiv.org/abs/2004.04562}{{\tt
  2004.04562}}.

\bibitem{Giombi:2023zte}
S.~Giombi, S.~Komatsu, B.~Offertaler, and J.~Shan, ``{Boundary
  reparametrizations and six-point functions on the AdS$_{2}$ string},'' {\em
  JHEP} {\bf 08} (2024) 196, \href{https://arxiv.org/abs/2308.10775}{{\tt
  2308.10775}}.

\bibitem{Rastelli:2016nze}
L.~Rastelli and X.~Zhou, ``{Mellin amplitudes for $AdS_5\times S^5$},'' {\em
  Phys. Rev. Lett.} {\bf 118} (2017), no.~9 091602,
  \href{https://arxiv.org/abs/1608.06624}{{\tt 1608.06624}}.

\bibitem{Rastelli:2017udc}
L.~Rastelli and X.~Zhou, ``{How to Succeed at Holographic Correlators Without
  Really Trying},'' {\em JHEP} {\bf 04} (2018) 014,
  \href{https://arxiv.org/abs/1710.05923}{{\tt 1710.05923}}.

\bibitem{Bissi:2022mrs}
A.~Bissi, A.~Sinha, and X.~Zhou, ``{Selected topics in analytic conformal
  bootstrap: A guided journey},'' {\em Phys. Rept.} {\bf 991} (2022) 1--89,
  \href{https://arxiv.org/abs/2202.08475}{{\tt 2202.08475}}.

\bibitem{Ferrero:2021bsb}
P.~Ferrero and C.~Meneghelli, ``{Bootstrapping the half-BPS line defect CFT in
  N=4 supersymmetric Yang-Mills theory at strong coupling},'' {\em Phys. Rev.
  D} {\bf 104} (2021), no.~8 L081703,
  \href{https://arxiv.org/abs/2103.10440}{{\tt 2103.10440}}.

\bibitem{Barrat:2021yvp}
J.~Barrat, A.~Gimenez-Grau, and P.~Liendo, ``{Bootstrapping holographic defect
  correlators in $ \mathcal{N} $ = 4 super Yang-Mills},'' {\em JHEP} {\bf 04}
  (2022) 093, \href{https://arxiv.org/abs/2108.13432}{{\tt 2108.13432}}.

\bibitem{Meneghelli:2022gps}
C.~Meneghelli and M.~Tr\'epanier, ``{Bootstrapping string dynamics in the 6d
  \ensuremath{\mathscr{N}} = (2, 0) theories},'' {\em JHEP} {\bf 07} (2023)
  165, \href{https://arxiv.org/abs/2212.05020}{{\tt 2212.05020}}.

\bibitem{Chen:2023oax}
J.~Chen and X.~Zhou, ``{Aspects of higher-point functions in BCFT$_{d}$},''
  {\em JHEP} {\bf 09} (2023) 204, \href{https://arxiv.org/abs/2304.11799}{{\tt
  2304.11799}}.

\bibitem{Gimenez-Grau:2023fcy}
A.~Gimenez-Grau, ``{The Witten Diagram Bootstrap for Holographic Defects},''
  \href{https://arxiv.org/abs/2306.11896}{{\tt 2306.11896}}.

\bibitem{Ferrero:2023znz}
P.~Ferrero and C.~Meneghelli, ``{Unmixing the Wilson line defect CFT. Part I.
  Spectrum and kinematics},'' {\em JHEP} {\bf 05} (2024) 090,
  \href{https://arxiv.org/abs/2312.12550}{{\tt 2312.12550}}.

\bibitem{Ferrero:2023gnu}
P.~Ferrero and C.~Meneghelli, ``{Unmixing the Wilson line defect CFT. Part II.
  Analytic bootstrap},'' {\em JHEP} {\bf 06} (2024) 010,
  \href{https://arxiv.org/abs/2312.12551}{{\tt 2312.12551}}.

\bibitem{Chen:2023yvw}
J.~Chen, A.~Gimenez-Grau, and X.~Zhou, ``{Defect two-point functions in 6D
  (2,0) theories},'' {\em Phys. Rev. D} {\bf 109} (2024), no.~6 L061903,
  \href{https://arxiv.org/abs/2310.19230}{{\tt 2310.19230}}.

\bibitem{Zhou:2024ekb}
X.~Zhou, ``{Correlators of N=4 Supersymmetric Yang-Mills Theory on Real
  Projective Space at Strong Coupling},'' {\em Phys. Rev. Lett.} {\bf 133}
  (2024), no.~20 201602, \href{https://arxiv.org/abs/2408.04926}{{\tt
  2408.04926}}.

\bibitem{Chen:2024orp}
J.~Chen, A.~Gimenez-Grau, H.~Paul, and X.~Zhou, ``{Unitarity method for
  holographic defects},'' {\em Phys. Rev. D} {\bf 111} (2025), no.~4 L041703,
  \href{https://arxiv.org/abs/2406.13287}{{\tt 2406.13287}}.

\bibitem{Chen:2025yxg}
J.~Chen, Y.~Jiang, and X.~Zhou, ``{Giant Graviton Correlators as Defect
  Systems},'' \href{https://arxiv.org/abs/2503.22987}{{\tt 2503.22987}}.

\bibitem{Beem:2013sza}
C.~Beem, M.~Lemos, P.~Liendo, W.~Peelaers, L.~Rastelli, and B.~C. van Rees,
  ``{Infinite Chiral Symmetry in Four Dimensions},'' {\em Commun. Math. Phys.}
  {\bf 336} (2015), no.~3 1359--1433,
  \href{https://arxiv.org/abs/1312.5344}{{\tt 1312.5344}}.

\bibitem{Beem:2014kka}
C.~Beem, L.~Rastelli, and B.~C. van Rees, ``{$ \mathcal{W} $ symmetry in six
  dimensions},'' {\em JHEP} {\bf 05} (2015) 017,
  \href{https://arxiv.org/abs/1404.1079}{{\tt 1404.1079}}.

\bibitem{Chester:2014mea}
S.~M. Chester, J.~Lee, S.~S. Pufu, and R.~Yacoby, ``{Exact Correlators of BPS
  Operators from the 3d Superconformal Bootstrap},'' {\em JHEP} {\bf 03} (2015)
  130, \href{https://arxiv.org/abs/1412.0334}{{\tt 1412.0334}}.

\bibitem{Beem:2016cbd}
C.~Beem, W.~Peelaers, and L.~Rastelli, ``{Deformation quantization and
  superconformal symmetry in three dimensions},'' {\em Commun. Math. Phys.}
  {\bf 354} (2017), no.~1 345--392,
  \href{https://arxiv.org/abs/1601.05378}{{\tt 1601.05378}}.

\bibitem{Liendo:2015ofa}
P.~Liendo, I.~Ramirez, and J.~Seo, ``{Stress-tensor OPE in $ \mathcal{N}=2 $
  superconformal theories},'' {\em JHEP} {\bf 02} (2016) 019,
  \href{https://arxiv.org/abs/1509.00033}{{\tt 1509.00033}}.

\bibitem{Lemos:2015orc}
M.~Lemos and P.~Liendo, ``{$\mathcal{N}=2$ central charge bounds from $2d$
  chiral algebras},'' {\em JHEP} {\bf 04} (2016) 004,
  \href{https://arxiv.org/abs/1511.07449}{{\tt 1511.07449}}.

\bibitem{Beem:2018duj}
C.~Beem, ``{Flavor Symmetries and Unitarity Bounds in ${\mathcal N}=2$
  Superconformal Field Theories},'' {\em Phys. Rev. Lett.} {\bf 122} (2019),
  no.~24 241603, \href{https://arxiv.org/abs/1812.06099}{{\tt 1812.06099}}.

\bibitem{Beem:2013qxa}
C.~Beem, L.~Rastelli, and B.~C. van Rees, ``{The $\mathcal N=4$ Superconformal
  Bootstrap},'' {\em Phys. Rev. Lett.} {\bf 111} (2013) 071601,
  \href{https://arxiv.org/abs/1304.1803}{{\tt 1304.1803}}.

\bibitem{Beem:2014zpa}
C.~Beem, M.~Lemos, P.~Liendo, L.~Rastelli, and B.~C. van Rees, ``{The $
  \mathcal{N}=2 $ superconformal bootstrap},'' {\em JHEP} {\bf 03} (2016) 183,
  \href{https://arxiv.org/abs/1412.7541}{{\tt 1412.7541}}.

\bibitem{Beem:2015aoa}
C.~Beem, M.~Lemos, L.~Rastelli, and B.~C. van Rees, ``{The (2, 0)
  superconformal bootstrap},'' {\em Phys. Rev. D} {\bf 93} (2016), no.~2
  025016, \href{https://arxiv.org/abs/1507.05637}{{\tt 1507.05637}}.

\bibitem{Lemos:2016xke}
M.~Lemos, P.~Liendo, C.~Meneghelli, and V.~Mitev, ``{Bootstrapping
  $\mathcal{N}=3$ superconformal theories},'' {\em JHEP} {\bf 04} (2017) 032,
  \href{https://arxiv.org/abs/1612.01536}{{\tt 1612.01536}}.

\bibitem{Beem:2016wfs}
C.~Beem, L.~Rastelli, and B.~C. van Rees, ``{More ${\mathcal N}=4$
  superconformal bootstrap},'' {\em Phys. Rev. D} {\bf 96} (2017), no.~4
  046014, \href{https://arxiv.org/abs/1612.02363}{{\tt 1612.02363}}.

\bibitem{Gimenez-Grau:2020jrx}
A.~Gimenez-Grau and P.~Liendo, ``{Bootstrapping Coulomb and Higgs branch
  operators},'' {\em JHEP} {\bf 01} (2021) 175,
  \href{https://arxiv.org/abs/2006.01847}{{\tt 2006.01847}}.

\bibitem{Bissi:2020jve}
A.~Bissi, A.~Manenti, and A.~Vichi, ``{Bootstrapping mixed correlators in $
  \mathcal{N} $ = 4 super Yang-Mills},'' {\em JHEP} {\bf 05} (2021) 111,
  \href{https://arxiv.org/abs/2010.15126}{{\tt 2010.15126}}.

\bibitem{Chester:2018dga}
S.~M. Chester and E.~Perlmutter, ``{M-Theory Reconstruction from (2,0) CFT and
  the Chiral Algebra Conjecture},'' {\em JHEP} {\bf 08} (2018) 116,
  \href{https://arxiv.org/abs/1805.00892}{{\tt 1805.00892}}.

\bibitem{Behan:2021pzk}
C.~Behan, P.~Ferrero, and X.~Zhou, ``{More on holographic correlators: Twisted
  and dimensionally reduced structures},'' {\em JHEP} {\bf 04} (2021) 008,
  \href{https://arxiv.org/abs/2101.04114}{{\tt 2101.04114}}.

\bibitem{Arutyunov:2002ff}
G.~Arutyunov and E.~Sokatchev, ``{Implications of superconformal symmetry for
  interacting (2,0) tensor multiplets},'' {\em Nucl. Phys. B} {\bf 635} (2002)
  3--32, \href{https://arxiv.org/abs/hep-th/0201145}{{\tt hep-th/0201145}}.

\bibitem{Heslop:2004du}
P.~J. Heslop, ``{Aspects of superconformal field theories in six dimensions},''
  {\em JHEP} {\bf 07} (2004) 056,
  \href{https://arxiv.org/abs/hep-th/0405245}{{\tt hep-th/0405245}}.

\bibitem{Rastelli:2017ymc}
L.~Rastelli and X.~Zhou, ``{Holographic Four-Point Functions in the (2, 0)
  Theory},'' {\em JHEP} {\bf 06} (2018) 087,
  \href{https://arxiv.org/abs/1712.02788}{{\tt 1712.02788}}.

\bibitem{Zhou:2017zaw}
X.~Zhou, ``{On Superconformal Four-Point Mellin Amplitudes in Dimension
  $d>2$},'' {\em JHEP} {\bf 08} (2018) 187,
  \href{https://arxiv.org/abs/1712.02800}{{\tt 1712.02800}}.

\bibitem{Heslop:2017sco}
P.~Heslop and A.~E. Lipstein, ``{M-theory Beyond The Supergravity
  Approximation},'' {\em JHEP} {\bf 02} (2018) 004,
  \href{https://arxiv.org/abs/1712.08570}{{\tt 1712.08570}}.

\bibitem{Abl:2019jhh}
T.~Abl, P.~Heslop, and A.~E. Lipstein, ``{Recursion relations for anomalous
  dimensions in the 6d $(2, 0)$ theory},'' {\em JHEP} {\bf 04} (2019) 038,
  \href{https://arxiv.org/abs/1902.00463}{{\tt 1902.00463}}.

\bibitem{Alday:2020lbp}
L.~F. Alday and X.~Zhou, ``{All Tree-Level Correlators for M-theory on $AdS_7
  \times S^4$},'' {\em Phys. Rev. Lett.} {\bf 125} (2020), no.~13 131604,
  \href{https://arxiv.org/abs/2006.06653}{{\tt 2006.06653}}.

\bibitem{Lemos:2021azv}
M.~Lemos, B.~C. van Rees, and X.~Zhao, ``{Regge trajectories for the (2, 0)
  theories},'' {\em JHEP} {\bf 01} (2022) 022,
  \href{https://arxiv.org/abs/2105.13361}{{\tt 2105.13361}}.

\bibitem{Kantor:2022epi}
G.~K\'antor, V.~Niarchos, C.~Papageorgakis, and P.~Richmond, ``{6D (2,0)
  bootstrap with the soft-actor-critic algorithm},'' {\em Phys. Rev. D} {\bf
  107} (2023), no.~2 025005, \href{https://arxiv.org/abs/2209.02801}{{\tt
  2209.02801}}.

\bibitem{Headrick:2015gba}
M.~Headrick, A.~Maloney, E.~Perlmutter, and I.~G. Zadeh, ``{R\'enyi entropies,
  the analytic bootstrap, and 3D quantum gravity at higher genus},'' {\em JHEP}
  {\bf 07} (2015) 059, \href{https://arxiv.org/abs/1503.07111}{{\tt
  1503.07111}}.

\bibitem{Woolley:2025qbd}
M.~Woolley, ``{The $\mathcal{W}$-algebra bootstrap of 6d $\mathcal{N}=(2,0)$
  theories},'' \href{https://arxiv.org/abs/2506.08094}{{\tt 2506.08094}}.

\bibitem{Lemos:2020pqv}
M.~Lemos, ``{Lectures on chiral algebras of $\mathcal{N} \geqslant 2$
  superconformal field theories},''
  \href{https://arxiv.org/abs/2006.13892}{{\tt 2006.13892}}.

\bibitem{watts1997}
G.~M.~T. Watts, ``$\mathcal{W}$-algebras and their representations,'' in {\em
  Conformal Field Theories and Integrable Models} (Z.~Horv{\'a}th and L.~Palla,
  eds.), (Berlin, Heidelberg), pp.~55--84, Springer Berlin Heidelberg, 1997.

\bibitem{Bouwknegt:1992wg}
P.~Bouwknegt and K.~Schoutens, ``{W symmetry in conformal field theory},'' {\em
  Phys. Rept.} {\bf 223} (1993) 183--276,
  \href{https://arxiv.org/abs/hep-th/9210010}{{\tt hep-th/9210010}}.

\bibitem{Fateev:1987zh}
V.~A. Fateev and S.~L. Lukyanov, ``{The Models of Two-Dimensional Conformal
  Quantum Field Theory with Z(n) Symmetry},'' {\em Int. J. Mod. Phys. A} {\bf
  3} (1988) 507.

\bibitem{Rastelli:2017ecj}
L.~Rastelli and X.~Zhou, ``{The Mellin Formalism for Boundary CFT$_d$},'' {\em
  JHEP} {\bf 10} (2017) 146, \href{https://arxiv.org/abs/1705.05362}{{\tt
  1705.05362}}.

\bibitem{Corrado:1999pi}
R.~Corrado, B.~Florea, and R.~McNees, ``{Correlation functions of operators and
  Wilson surfaces in the d = 6, (0,2) theory in the large N limit},'' {\em
  Phys. Rev. D} {\bf 60} (1999) 085011,
  \href{https://arxiv.org/abs/hep-th/9902153}{{\tt hep-th/9902153}}.

\bibitem{Bastianelli:1999en}
F.~Bastianelli and R.~Zucchini, ``{Three point functions of chiral primary
  operators in d = 3, N=8 and d = 6, N=(2,0) SCFT at large N},'' {\em Phys.
  Lett. B} {\bf 467} (1999) 61--66,
  \href{https://arxiv.org/abs/hep-th/9907047}{{\tt hep-th/9907047}}.

\bibitem{Beccaria:2024gkq}
M.~Beccaria, L.~Casarin, and A.~A. Tseytlin, ``{Semiclassical quantization of
  M5 brane probes wrapped on AdS$_{3}$\texttimes{} S$^{3}$ and defect
  anomalies},'' {\em JHEP} {\bf 01} (2025) 088,
  \href{https://arxiv.org/abs/2411.11626}{{\tt 2411.11626}}.

\bibitem{Goncalves:2025jcg}
V.~Goncalves, M.~Nocchi, and X.~Zhou, ``{Dissecting supergraviton six-point
  function with lightcone limits and chiral algebra},''
  \href{https://arxiv.org/abs/2502.10269}{{\tt 2502.10269}}.

\bibitem{Alday:2024srr}
L.~F. Alday and X.~Zhou, ``{Flat-space limit of defect correlators and stringy
  AdS form factors},'' {\em JHEP} {\bf 03} (2025) 182,
  \href{https://arxiv.org/abs/2411.04378}{{\tt 2411.04378}}.

\end{thebibliography}\endgroup
%%%%%%%%%%%%%%%%%%%%%%%%%%%%%%%%%%%%%%%%%%%%%%%%%%%%%%%%%%%%%%%%%%%%%%%%%%%%%%%%%%%%%%%%%%%%%%
\end{document}